\documentclass[
    ,final            
  ]
  {aipproc}

\layoutstyle{8x11double}

\begin{document}

\title{PLANETARY NEBULAE AS A CHEMICAL EVOLUTION TOOL: 
ABUNDANCE GRADIENTS}



\author{W. J. Maciel, L. G. Lago, R. D. D. Costa}{
  address={IAG/USP, S\~ao Paulo, Brazil}
}

\begin{abstract}
We have studied the time variation of the radial abundance
gradients using samples of planetary nebulae, open clusters,
cepheids and other young objects. Based on the analysis of
O/H and S/H abundances for planetary nebulae and  
metallicities of the remaining objects, we concluded that 
the gradients have been flattening out in the last
8 Gyr with an average rate of the order of 0.005 -- 0.010 dex kpc$^{-1}$ 
Gyr$^{-1}$. We have estimated the errors involved in the
determination of the gradients, and concluded that
the existence of systematic abundance variations is more likely
than a simple statistical dispersion around a mean value.
\end{abstract}

\maketitle

\section{1. Introduction}

Radial abundance gradients in the Milky Way disk are among the main 
constraints of models of the chemical evolution of the Galaxy. The study 
of the gradients comprises the determination of their magnitudes along the 
disk, space variations and their time evolution (see for example Henry \& 
Worthey 1999, Maciel \& Costa  2003). Probably the most interesting property 
of the gradients is their time evolution, which is a distinctive constraint 
of  recent chemical evolution models. Maciel et al. (2003) suggested that 
the O/H gradient has been flattening during the last few Gyr, on the basis 
of a large sample of planetary nebulae (PN) for which accurate abundances 
are available, and for which the ages of the progenitor stars have been 
individually estimated. This work has been recently extended (Maciel et al. 
2005) to include the S/H ratio in planetary nebulae, [Fe/H] metallicities
from open clusters and cepheid variables, as well as some young objects, 
such as OB associations and HII regions.

In this work, we review the main characteristics of the work by Maciel et 
al. (2005) and analyze the uncertainties involved in the determination of 
the gradients. In particular, we investigate whether the derived 
uncertainties support either a systematic variation of the abundances with 
the galactocentric distance, as assumed by our work, or simply a dispersion 
of the abundances around some average value.

\section{2. Abundance Gradients}

The main results for the time variation of the gradients as derived from 
planetary nebulae, open clusters, and cepheids are shown in tables~1 and 2. 
Adopting average linear gradients, which can be taken as representative 
of the whole galactic disk, the abundances can be written in the form

\begin{equation}
y = a + b\ R
\end{equation}

\noindent
where $y = \log$(O/H) + 12 or $y = \log$(S/H) + 12 for PN, HII regions and 
OB stars, and $y = $ [Fe/H] for open clusters and cepheids. For planetary 
nebulae, we have taken into account both O/H and S/H determinations and 
evaluated the gradient in the galactic disk according to the ages of the 
progenitor stars. For comparison purposes, we can also derive the [Fe/H]
metallicities from the O/H abundances, on the basis of a [Fe/H]  $\times$ 
O/H correlation derived for disk stars (see Maciel 2002 and Maciel et al. 
2005 for details). The ages follow from the age-metallicity relation by 
Edvardsson et al. (1993), which also depends on the galactocentric distance. 
In this way, we can divide the sample of PN into different age groups, each 
one having a characteristic gradient.

Table 1 shows representative examples of 3 age groups for O/H and 2 age 
groups for S/H. The table gives the gradient $b$ (dex/kpc) as defined by 
equation (1). All gradients in this paper have been calculated assuming 
$R_0 = 8.0$ kpc for the galactocentric distance of the LSR. For detailed 
references on the PN data the reader is referred to Maciel et al. (2003, 
2005). It should be mentioned that the PN age groups shown in Table~1 are 
typical groups, arbitrarily defined. In fact, we have extended this 
procedure by taking into account a variety of definitions of the age groups, 
with similar results.

\begin{table}
\begin{tabular}{lrr}
\hline
  \tablehead{1}{r}{b}{Group}
  & \tablehead{1}{r}{b}{Age (Gyr)}
  & \tablehead{1}{r}{b}{$b$ (dex/kpc)}\\
\hline
O/H  &         &                     \\
I    &  $0-4$  &  $-0.047\pm 0.007$  \\
II   &  $4-5$  &  $-0.089\pm 0.003$  \\
III  &  $5-8$  &  $-0.094\pm 0.010$  \\
     &         &                     \\
S/H  &         &                     \\
I    &  $0-4$  & $-0.080\pm 0.014$   \\
II   &  $4-8$  & $-0.113\pm 0.011$   \\
\hline
\end{tabular}
\caption{Planetary nebulae}
\label{tab:1}
\end{table}

\begin{table}
\begin{tabular}{lrr}
\hline
  \tablehead{1}{r}{b}{Group}
  & \tablehead{1}{r}{b}{Age (Gyr)}
  & \tablehead{1}{r}{b}{$b$ (dex/kpc)}\\
\hline
[Fe/H] &         &     \\
OC - Friel &         &     \\
I    &  $0-2$   &  $-0.023\pm 0.019$  \\
II   &  $2-5$   &  $-0.058\pm 0.014$  \\
III  &  $5-8$   &  $-0.104\pm 0.035$  \\
     &          &                     \\
OC - Chen &          &     \\
I    &  $0-0.8$ &  $-0.024\pm 0.012$  \\
II   &  $0.8-2$ &  $-0.067\pm 0.018$  \\
III  &  $2-8$   &  $-0.084\pm 0.020$  \\
     &          &                     \\
Cepheids   &  $0-700$ Myr   & $-0.054\pm 0.003$   \\
\hline
\end{tabular}
\caption{Open clusters and cepheids}
\label{tab:2}
\end{table}

For open clusters we have adopted a similar procedure, except that we have
used two different samples: (i) the smaller although homogeneous sample by
Friel et al. (2002) and (ii) the larger compilation by Chen et al. (2003). 
As we will see, there are no important differences in the results from
both samples. We have also divided the samples into age groups, of which
the ones shown in Table~2 are representative. In this case, the given
slopes $b$ refer directly to the [Fe/H] gradients.

Also shown in Table~2 is the derived gradient from cepheid variables.
Here we have adopted the recent detailed work by Andrievsky and coworkers
(see Andrievsky et al. 2002 and 2004 for the first and last papers in the 
series), and considered the [Fe/H] abundances, which are the best 
determined. The ages given in Table 2 have been derived by us on the 
basis of metallicity dependent period-luminosity relationships.

Finally, we have also included some data on young objects, so that we can 
have a complete view of the time variation of the abundance gradients. We 
have considered HII regions, for which we have adopted an average gradient
from the recent literature, namely $b = d\log({\rm O/H})/dR = -0.055
\pm 0.015$\  (see for example Henry \& Worthey 1999, Deharveng et al. 2000, 
Pilyugin et al. 2003). Also, a sample of OB stars and associations recently 
studied by Daflon \& Cunha (2004) have been taken into account, with a
gradient $b = d\log({\rm O/H})/dR = -0.031 \pm 0.012$\ . These are very 
young objects indeed, and from the point of view of the present work, 
their ages can be considered as essentially zero, in the case of HII 
regions, and about $0-50$ Myr for OB stars. In both cases we can also 
convert the O/H gradients into [Fe/H] using the same relation adopted for 
planetary nebulae.

As an illustration, Figure~1 shows the derived gradients, converted to
[Fe/H] gradients, for all objects considered in this work. It can be seen
that the gradients are clearly flattening out in the last 6 to 8 Gyr,
assuming an age of 13.6 Gyr for the galactic disk.

\begin{figure}
  \includegraphics[angle=-90.0,width=7.75cm]{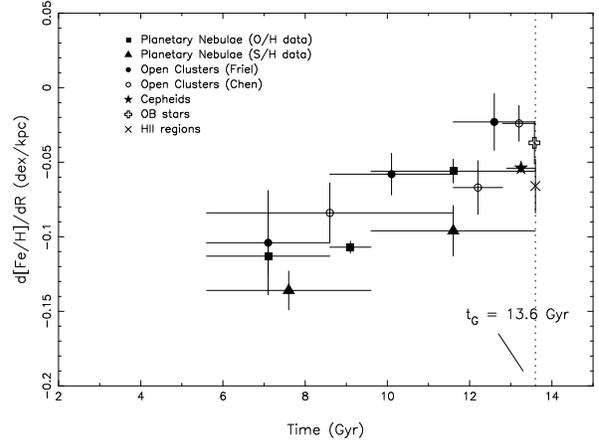}
  \caption{Time variation of the [Fe/H] gradient.}
\end{figure}

\section{3. Statistical Analysis}

\subsection{3.1 Correlation coefficients and sample size}

The simplest way to analyze the uncertainties of the linear fits is to 
consider the derived correlation coefficients. We have $0.64 < \vert r\vert 
< 0.94$ for planetary nebulae, $0.33 < \vert r\vert < 0.77$ for open clusters 
(Friel et al. sample), $0.22 < \vert r\vert < 0.71$ for open clusters 
(Chen et al. sample), and $\vert r\vert = 0.82$ for cepheid variables. 
This suggests that the PN and cepheid gradients are well determined, 
while the analysis of the open cluster data is more complex. In fact, 
the correlation coefficient depends on the age group considered, being 
larger for groups II and III and smaller for the youngest group. Since 
Group~I is the largest in both samples, there may be some contamination 
from older objects, or some effect derived from the space distribution 
of the clusters. In particular, young clusters seem to concentrate in the 
inner parts of the galactic disk, so that their distribution presents a 
more limited range of galactocentric distances, contributing to a 
flatter gradient.

The derived correlation coefficients along with the sample sizes can be 
used in order to compare these values with the probability distribution 
of a parent population which is completely uncorrelated. In this way we 
can have an indication on whether or not it is probable that our data 
points could be used to represent a sample derived from an uncorrelated 
population. Naturally, if this probability is small, we can conclude that 
our data points represent a sample derived from a parent population in which
the quantities involved, that is, abundances and galactocentric distances, 
are correlated.

We have then estimated the probability $P(r,n)$ that a given sample 
containing $n$ objects, for which a linear correlation coefficient $r$ was 
derived, could have come from an uncorrelated parent population, assuming 
gaussian distributions. The results are shown in Table~3 for PN, open 
clusters and cepheid variables.

It can be seen that the probability $P(r,n)$ is exceedingly small in all 
cases, in fact several orders of magnitude lower than the indicated value 
of 0.001. Again, the open clusters are exceptions, especially the youngest 
groups in both samples, but even in this case the probability $P(r,n)$ is 
still small. In other words, the probability of obtaining a large correlation 
coefficient as shown in Table~3, on the basis of samples of the given sizes 
from an uncorrelated population is generally very small.

\begin{table}
\begin{tabular}{lrrrr}
\hline
  & \tablehead{1}{r}{b}{Group}
  & \tablehead{1}{r}{b}{$\mid r\mid$}
  & \tablehead{1}{r}{b}{$n$}
  & \tablehead{1}{r}{b}{$P(r,n)$}   \\
\hline
O/H -- PN   & I   & 0.64  &  66 & $<$0.001 \\
            & II  & 0.94  &  99 & $<$0.001 \\
            & III & 0.75  &  69 & $<$0.001 \\
            &     &       &     &          \\
S/H -- PN   & I   & 0.66  &  44 & $<$0.001 \\
            &  II & 0.76  &  72 & $<$0.001 \\
            &     &       &     &          \\
OC -- Friel & I   & 0.33  &  15 & 0.230    \\
            & II  & 0.77  &  14 & 0.001    \\ 
            & III & 0.73  &  10 & 0.017    \\
            &     &       &     &          \\
OC -- Chen  & I   & 0.22  &  80 & 0.051    \\
            & II  & 0.69  &  18 & 0.002    \\
            & III & 0.71  &  20 & $<$0.001 \\
            &     &       &     &          \\
Cepheids    &     & 0.82  & 127 & $<$0.001 \\
\hline
\end{tabular}
\caption{Probability of uncorrelated parent population $P(r,n)$}
\label{tab:3}
\end{table}

\subsection{3.2 Average uncertainties}

It is interesting to compare the uncertainties associated with the linear 
fits with those associated with a simple dispersion around some average value, 
that is, assuming no variations of the abundances with the galactocentric 
distance.  Since the observational abundance uncertainties are generally 
taken to be in the range 0.1 -- 0.2 dex, we find that the uncertainties of 
the linear correlations  are usually much lower than the uncertainties 
obtained assuming no systematic variations with $R$. 

\begin{table}
\begin{tabular}{lrr}
  \tablehead{1}{r}{b}{PN}
  & \tablehead{1}{r}{b}{gradient}
  & \tablehead{1}{r}{b}{no gradient}\\
{\bf O/H} &     &                   \\
I    & 0.09     & 0.19              \\
II   & 0.04     & 0.18              \\
III  & 0.12     & 0.26              \\
{\bf S/H} &     &                   \\
I    & 0.19     & 0.30              \\
II   & 0.14     & 0.32              \\
\end{tabular}
\label{tab:a}
\end{table}

For planetary nebulae this is very clear, which seems  
to be a very good argument in favour of the existence of systematic 
abundance variations.

For open clusters the comparison is not so clear. In the case of the large 
sample by Chen et al. (2003) the uncertainties derived assuming systematic 
variations are lower or similar to those derived with no systematic 
variations, but the data by Friel shows the opposite behaviour. In the 
case of Group~I, especially the Friel sample, we have seen that the
gradients are not very well defined, which helps to explain the large
uncertainties given above.

\begin{table}
\begin{tabular}{lrr}
  \tablehead{1}{r}{b}{OC}
  & \tablehead{1}{r}{b}{gradient}
  & \tablehead{1}{r}{b}{no gradient}\\
{\bf Friel} &    &                  \\
I     & 0.25     & 0.09             \\
II    & 0.20     & 0.16             \\
III   & 0.46     & 0.20             \\
{\bf Chen} &     &                  \\
I     & 0.15     & 0.15             \\
II    & 0.24     & 0.21             \\
III   & 0.29     & 0.31             \\
\end{tabular}
\label{tab:b}
\end{table}

For Cepheids, the uncertainties are usually lower than in all other cases, 
as expected. Here, the uncertainties assuming a gradient are much lower then 
otherwise (0.04 dex for the case with gradients compared to 0.13 dex otherwise), 
which reinforces our conclusions about the reality of the gradients.

\subsection{3.3 $\chi^2$ analysis}

Assuming that the abundances are distributed around an average value, 
characterized by the mean $\mu$ and standard deviation $\sigma$, we can 
estimate the reduced $\chi^2_\nu$ \ and the probability $P(\chi^2,\nu)$ 
that a random sample of data points would yield a value of $\chi^2_\nu$ 
\ as large as or larger than the observed if the parent distribution, 
assumed gaussian, were equal to the assumed distribution. If the 
probability is approximately $P(\chi^2,\nu) \simeq 1$, the assumed 
distribution describes the spread of data points well. If the probability 
is small, either the assumed distribution is not a good estimate of the 
parent distribution or the data are not a representative sample.

\begin{table}
\begin{tabular}{lrr}
\hline
  \tablehead{1}{r}{b}{Group}
  & \tablehead{1}{r}{b}{\ \ \ $\chi^2_\nu$\ \ [$P(\chi^2,\nu]$\quad }
  & \tablehead{1}{r}{b}{$\chi^2_\nu$\ \ [$P(\chi^2,\nu]$\quad \quad }  \\
  &   average \quad \quad  &   linear \quad \quad \quad             \\
\hline
PN     &                  &                             \\
O/H    &                  &                             \\
 I     &  1.69\ \ [0.06]  &  0.98--0.55\ \ [0.52--1.00] \\
 II    &  1.15\ \ [0.33]  &  0.17--0.10\ \ [1.00--1.00] \\
 III   &  1.26\ \ [0.23]  &  1.31--0.74\ \ [0.05--1.00] \\
S/H    &                  &                             \\
I      &  2.09\ \ [0.01]  &  1.28--0.82\ \ [0.10--0.79] \\
II     &  0.98\ \ [0.49]  &  1.09--0.70\ \ [0.28--0.97] \\
       &                  &                             \\
OC     &                  &                             \\
Friel  &                  &                             \\
I      &  0.17\ \ [0.96)  &  0.20--0.13\ \ [1.00--1.00] \\
II     &  0.67\ \ [0.68)  &  0.29--0.18\ \ [0.99--1.00] \\
III    &  1.28\ \ [0.27)  &  0.50--0.32\ \ [0.85--0.96] \\
Chen   &                  &                             \\
I      &  3.82\ \ [0.00]  &  0.56--0.36\ \ [1.00--1.00] \\
II     &  0.33\ \ [0.97]  &  0.60--0.38\ \ [0.89--0.99] \\
III    &  0.84\ \ [0.62]  &  1.26--0.80\ \ [0.21--0.70] \\
       &                  &                             \\
Cepheids &                &                             \\
       &  1.96\ \ [0.04]  &  0.60--0.27\ \ [1.00--1.00] \\
\hline
\end{tabular}
\caption{$\chi^2_\nu$ and probability $P(\chi^2,\nu)$}
\label{tab:4}
\end{table}

Column~2 of Table~4 shows the estimated values of $\chi^2_\nu$\ and 
$P(\chi^2,\nu)$\ [within brackets] assuming average values, that is,
no linear variations. The results for PN show that the probability is very 
low in all cases, so that the data points are probably not distributed 
according to a gaussian distribution around some average value. However, 
it is interesting to note that, if we restrain the galactocentric distances 
to a smaller range, such as from $R = 6$ kpc to 8 kpc, or $R = 8$ kpc to 
10 kpc, the probability $P(\chi^2,\nu)$ increases, showing that, for a 
given galactocentric bin, the abundances show a better agreement with the 
gaussian distribution around some average value. 

For the open clusters, the table shows a generally better agreement with 
the gaussian distribution around a mean value, both for the Friel and Chen 
samples, in agreement with our conclusions in sect. 3.2. However, for 
cepheid variables we have the same results as for the PN, that is, the 
cepheid data are apparently not consistent with a gaussian distribution 
around a mean value. 

We can also estimate $P(\chi^2,\nu)$ in each case taking into account the 
derived linear correlations which are displayed in Tables~1 and 2. Here we 
have $\nu = n - 2$ for the number of degrees of freedom, so that we can 
estimate $\chi^2_\nu$ and $P(\chi^2,\nu)$\ provided we have a reliable 
estimate of the uncertainty of the data. For planetary nebulae, recent 
discussions by Pottasch et al. (2005) of objects with ISO data suggest 
that the abundances of the beststudied elements are probably correct within 
20\%, which corresponds to 0.10 dex for oxygen. This is probably a lower 
limit for other nebulae for which no infrared data is available, so that 
their abundances depend more heavily on ionization correction factors.
We may then adopt $\sigma \simeq 0.15-0.20$ dex  for O/H  and 
$\sigma \simeq 0.20-0.25$ dex for S/H as realistic estimates for 
planetary nebulae. The latter can also be attributed to the open clusters,
in view of the heterogeneity of the data and the use of photometric 
abundances. For cepheid variables, which have the best determinations, 
an average uncertainty $\sigma \simeq 0.10-0.15$ seems appropriate. The 
results are shown in column~3 of Table~4, under the  heading \lq\lq 
linear\rq\rq. Again the probabiliy is given within brackets. We can see 
that in all cases the  $\chi^2_\nu$\ values are lower than the 
corresponding values for the averages, so that the probability 
$P(\chi^2,\nu)$\ is higher for the linear correlation than for the simple 
averages. In fact, these probabilities are very close to unity in most cases, 
especially if we consider the more realistic, higher 
uncertainties. It can  also be seen that for cepheid variables the 
probability given in column~3 is essentially unity, reinforcing our conclusion 
about systematic abundance variations with the galactocentric distance.

\begin{theacknowledgments}

This work has been partially supported by FAPESP,
CNPq and CAPES.

\end{theacknowledgments}



\bibliographystyle{aipproc}   


\IfFileExists{\jobname.bbl}{}
 {\typeout{}
  \typeout{******************************************}
  \typeout{** Please run "bibtex \jobname" to optain}
  \typeout{** the bibliography and then re-run LaTeX}
  \typeout{** twice to fix the references!}
  \typeout{******************************************}
  \typeout{}
 }

\end{document}